\documentstyle[prl,aps,epsf,floats]{revtex}
\begin{document}
\draft

\twocolumn[\hsize\textwidth\columnwidth\hsize\csname @twocolumnfalse\endcsname

\title{
Three-body scattering problem and 
two-electron tunneling in molecular wires
}
\author{
A.S. Alexandrov$^{1,2}$,  A.M. Bratkovsky$^1$, and 
P.E. Kornilovitch$^1$
}
\address{$^1$Hewlett-Packard Laboratories, 1501 Page Mill 
Road, MS 1L-12, Palo Alto, California 94304\\
$^2$Department of Physics, Loughborough University,
Loughborough LE11 3TU, United Kingdom}
\maketitle

\begin{abstract}
We solve the Lippmann-Schwinger equation describing elastic scattering of
 preformed pairs (e.g. bipolarons) off a short-range scattering center and
find the two-particle transmission through a thin potential barrier.
While the pair transmission is smaller than the single-electron transmission
in the strong-coupling limit, it is remarkably larger in the weak coupling
limit. We also calculate current-voltage characteristics of a molecule - barrier - molecule 
 junction. They show unusual temperature and voltage behavior
which are  experimentally verifiable at low temperatures.
\end{abstract}

\vskip2pc]

\narrowtext

Molecular-scale electronics is currently a very active area of research\cite
{mark98}. It is envisaged that linear conjugated molecules would be used as
the ``transmission lines'' in molecular circuitry\cite{lehn90,tour00} in
addition to active molecular elements discussed in the literature \cite
{mark98,pat99}. When a so-called ``molecular wire'' is short, the dominant
mechanism of transport is most likely a resonant tunneling through
electronic molecular states (see \cite{restunn,kb01} and references
therein). With increasing size of the wires one has to take into account
strong interaction between carriers and vibronic excitations of the
molecule, leading to self-trapping of electrons in polaronic states. The
formation of polarons (and charged solitons) in polyacetylene (PA)\ was
discussed theoretically in Refs. \cite{su80} and formation of bipolarons
(bound states of two polarons) in Ref.\cite{braz81}. Polarons in PA were
detected optically in Ref.\cite{feldblum82} and since then studied in great
detail. There is also an exceeding amount of evidence of the polaron and
bipolaron formation in conjugated polymers such as polyphenylene,
polypyrrole, polythiophene, polyphenylene sulfide \cite{bredas84}, Cs-doped
biphenyl \cite{ramsey90}, n-doped bithiophene \cite{steinmuller93},
polyphenylenevinylene(PPV)-based light emitting diodes \cite{swanson93}, and
other molecular systems. In many cases the doped polymers have
bipolaron-like charge states to yield, in particular, the enhanced nonlinear
optical properties\cite{spangler94}.

Many experimental data provide evidence for hopping transport of
(bi)polaronic carriers. However, at sufficiently low temperatures there
should be a crossover to the band motion of polarons, as suggested long ago 
\cite{hol,lan}, and bipolarons\cite{alemot,tru}. Indeed, due to recent
extraordinary improvements in preparation of ``plastic'' molecular
conductors, it became possible to measure their conductivity in a wide
interval of temperatures and observe the crossover in two-dimensional films
of organic conjugated molecules \cite{bat}. In one-dimensional (1D) wires
the band motion is expected to be strongly hindered by imperfections, and
those imperfections are likely to be intentionally introduced in the system
as functionalizing units\cite{reed00}. Moreover, the polarons in extended
molecular wires/units are expected to be bound into real space bipolarons
with lowering temperature. As it is known in the context of oxide
semiconductors the bipolaron formation may strongly affect transport
properties \cite{alemot,abcmr}.

In this Letter we study elastic scattering of carriers bound into real-space
pairs in one-dimensional organic and other conductors. We present an exact
analytical solution in the limit of slow pairs. We also find an unusual
temperature and voltage dependence of the tunneling conductance which may be
experimentally verified at low temperatures.

In mathematical terms, the scattering of pairs is a three-body problem with
the mass of the third particle taken to infinity. Let $\hat{U}(x_{1}-x_{2})$
be an attractive potential between the two moving particles and $\hat{V}%
(x_{1},x_{2})$ the repulsive external potential representing the barrier.
The starting point is the Lippmann-Schwinger equation \cite{lip,baz} for the
two-particle wave function $\Psi (k_{1},k_{2})$ in momentum representation,
which explicitly takes into account a boundary condition of the three-body
scattering problem. It can be written as 
\begin{equation}
\Psi =-i\gamma \hat{G}(E+i\gamma )\Phi ,  \label{one}
\end{equation}
where $\hat{G}(E+i\gamma )$ is the exact two-particle Green's function (GF)
in the external potential, $\Phi (k_{1},k_{2})$ is the wave function of a
free ($\hat{V}=0$) real-space pair in momentum representation, $\Phi
(k_{1},k_{2})=2\pi \delta (q-Q)\phi (k)$. Here $q=k_{1}+k_{2}$ is the
center-of-mass momentum, $k=(k_{1}-k_{2})/2$ is the relative momentum, $%
E=-\epsilon +Q^{2}/4<0$ is the pair total energy in the absence of the
external potential, and $\epsilon $ is its binding energy. The wave function 
$\phi (k)$ describes the internal structure of the pair. Hereafter we choose 
$\hbar =k_{B}=m_{1}=m_{2}=1,\gamma =+0$, and define $\hat{G}(E+i\gamma )\Phi 
$ as 
\begin{equation}
\hat{G}\Phi \equiv \int \int \frac{dk_{1}^{\prime }dk_{2}^{\prime }}{(2\pi
)^{2}}G(k_{1},k_{2}|k_{1}^{\prime },k_{2}^{\prime };E)\Phi (k_{1}^{\prime
},k_{2}^{\prime }),  \label{two}
\end{equation}
for any $\hat{G}$ and $\Phi $. Using the two identities, $\hat{G}=\hat{G}%
_{3}-\hat{G}_{3}\hat{U}\hat{G}$ and $\hat{G}=\hat{G}_{12}-\hat{G}_{12}\hat{V}%
\hat{G}$ and the Lippmann-Schwinger equation one readily derives the
equation for the Fourier component $T(k_{1},k_{2})$ of the product $\hat{U}%
\Psi $ \cite{baz} 
\begin{equation}
T(k_{1},k_{2})=(E-k^{2}-q^{2}/4)\Phi -\hat{T}_{12}\Delta \hat{G}_{3}T,
\label{three}
\end{equation}
where $\Delta \hat{G}_{3}=\hat{G}_{3}-\hat{G}_{0}$, $\hat{G}_{0}$ is the
two-particle GF in the absence of any interaction ($\hat{U}=\hat{V}=0$), $%
\hat{G}_{3}$ is the GF of noninteracting particles in the external field $%
\hat{V}$ (for $\hat{U}=0)$, and $\hat{G}_{12}$ is GF of two interacting
particles with no external field, $\hat{V}=0$. Here the scattering operator $%
\hat{T}_{12}$,defined by the relation $\hat{T}_{12}\hat{G}_{0}\equiv \hat{U}%
\hat{G}_{12}$, is expressed via the particle-particle scattering $t$-matrix
as 
\begin{equation}
T_{12}(k_{1},k_{2}|k_{1}^{\prime },k_{2}^{\prime };E)=2\pi t(k,k^{\prime
};E-q^{2}/4)\delta (q-q^{\prime }).  \label{four}
\end{equation}
The $t$-matrix satisfies the equation 
\begin{equation}
t(k,k^{\prime };E)=u(k-k^{\prime })-\int \frac{dp}{2\pi }\frac{%
u(k-p)t(p,k^{\prime };E)}{p^{2}-E-i\gamma },  \label{five}
\end{equation}
where $u(p)$ is the Fourier component of the attractive potential $\hat{U}%
(x_{1}-x_{2})$.

In many (in)organic semiconductors, the long-range Coulomb repulsion is
usually significantly reduced by the strong Fr\"{o}hlich interaction with
optical phonons \cite{ale}, so that a net (attractive) potential between
carriers is a short-range one, $\hat{U}(x_{1}-x_{2})=-\alpha \delta
(x_{1}-x_{2}),$ $\alpha >0$. Then Eq.(\ref{five}) is readily solved
resulting in the momenta-independent $t$-matrix 
\begin{equation}
t(k,k^{\prime },E)=-\frac{\alpha \sqrt{-E}}{\sqrt{-E}-\alpha /2}{,}
\label{six}
\end{equation}
which is valid for all energies provided that the square root is understood
as its principal value. The binding energy is $\epsilon =\alpha ^{2}/4,$ and
the normalized ground state wave function is $\phi (k)=2^{-1/2}\alpha
^{3/2}/(k^{2}+\epsilon )$. It is known that for a short-range inter-particle
interaction $T(k_{1},k_{2})$, Eq.(\ref{three}), is proportional to the
Fourier component of the center-of-mass wave function $\Omega (q)$, $%
T(k_{1},k_{2})=-2^{-1/2}\alpha ^{3/2}\Omega (q)$. Then the problem of
elastic pair scattering is reduced to a single integral equation for the
center-of-mass scattering amplitude $\Upsilon (q)$. Substituting $\Phi $ and 
$\hat{T}_{12}$ in Eq.~(3) one obtains 
\begin{equation}
\Omega (q)=2\pi \delta (q-Q)-\frac{\Upsilon (q)}{q^{2}/4-Q^{2}/4-i\gamma },
\label{seven}
\end{equation}
where $\Upsilon (q)$ satisfies 
\begin{equation}
\Upsilon (q)=W(q,Q)-\int \frac{dq^{\prime }}{2\pi }\frac{W(q,q^{\prime
})\Upsilon (q^{\prime })}{q^{\prime 2}/4-Q^{2}/4-i\gamma }.  \label{eight}
\end{equation}
The effective center-of-mass scattering potential $W(q,q^{\prime })$ is
determined using GF of two noninteracting particles in the external
potential ($\hat{U}=0$ but $\hat{V}\neq 0$) as 
\begin{equation}
W(q,q^{\prime })=\alpha \chi (q)\int \int \frac{dk_{2}dk_{2}^{\prime }}{%
(2\pi )^{2}}\Delta G_{3}(q-k_{2},k_{2}|q^{\prime }-k_{2}^{\prime
},k_{2}^{\prime };E),  \label{nine}
\end{equation}
where $\chi (q)=E-q^{2}/4+(\alpha /4)(q^{2}-4E)^{1/2}.$ In the following we
restrict our consideration to the scattering of slow pairs with $Q^{2}\ll
4\epsilon $. This condition allows us to replace $W(q,q^{\prime })$ with $%
W(0,0)\equiv W$ in all equations because the characteristic momenta $%
q,q^{\prime }\simeq Q$ are much smaller than $\sqrt{-E}$. Then the solution
of Eq.~(\ref{eight}) is given by $\Upsilon (q)=WQ/(Q+2iW)$ so that the pair
transmission probability is 
\begin{equation}
{\cal T}_{2}(Q)=1-\left| \frac{2\Upsilon (-Q)}{Q}\right| ^{2}=\frac{Q^{2}}{%
Q^{2}+4W^{2}}.  \label{eleven}
\end{equation}
In general, $G_{3}$, $W$, and ${\cal T}_{2}$ can be found only numerically.
However, in many applications the scattering potential is also a short-range
one, $\hat{V}(x_{1},x_{2})=\beta \lbrack \delta (x_{1})+\delta (x_{2})]$, so
that the full Hamiltonian takes the form 
\begin{equation}
H=-\frac{1}{2}\frac{\partial ^{2}}{\partial x_{1}^{2}}-\frac{1}{2}\frac{%
\partial ^{2}}{\partial x_{2}^{2}}-\alpha \delta (x_{1}-x_{2})+\beta \left[
\delta (x_{1})+\delta (x_{2})\right] .  \label{elevenone}
\end{equation}
This three-body problem was considered before in \cite{sch,jas} but no
general analytical solution was found. Here we present the analytical
solution in the limit of slow pairs. Consider the equation for the
two-particle $G_{3}$ 
\begin{eqnarray}
&&(k_{1}^{2}/2+k_{2}^{2}/2-E)G_{3}(k_{1},k_{2}|k_{1}^{\prime },k_{2}^{\prime
};E)+  \nonumber \\
&&\beta \int \frac{dp_{1}}{2\pi }G_{3}(p_{1},k_{2}|k_{1}^{\prime
},k_{2}^{\prime };E)+\beta \int \frac{dp_{2}}{2\pi }%
G_{3}(k_{1},p_{2}|k_{1}^{\prime },k_{2}^{\prime };E)  \nonumber \\
&=&(2\pi )^{2}\delta (k_{1}-k_{1}^{\prime })\delta (k_{2}-k_{2}^{\prime }),
\label{twelve}
\end{eqnarray}
which has a formal solution 
\begin{eqnarray}
G_{3}(k_{1},k_{2}| &&k_{1}^{\prime },k_{2}^{\prime
};E)=G_{0}(k_{1},k_{2}|k_{1}^{\prime },k_{2}^{\prime };E)  \nonumber \\
&&-\frac{D(k_{2}|k_{1}^{\prime },k_{2}^{\prime };E)+D(k_{1}|k_{2}^{\prime
},k_{1}^{\prime };E)}{k_{1}^{2}/2+k_{2}^{2}/2-E}.  \label{thirteen}
\end{eqnarray}
Here $G_{0}(k_{1},k_{2}|k_{1}^{\prime },k_{2}^{\prime };E)=(2\pi )^{2}\delta
(k_{1}-k_{1}^{\prime })\delta (k_{2}-k_{2}^{\prime
})(k_{1}^{2}/2+k_{2}^{2}/2-E)^{-1},$ and $D(k_{1}|k_{2}^{\prime
},k_{1}^{\prime };E)\equiv (2\pi )^{-1}\beta \int
dk_{2}G_{3}(k_{1},k_{2}|k_{1}^{\prime },k_{2}^{\prime };E)$ satisfies the
integral equation 
\begin{eqnarray}
&&D(k_{1}|k_{2}^{\prime },k_{1}^{\prime };E)~\left[ 1+\frac{\beta }{%
(k_{1}^{2}-2E)^{1/2}}\right] =  \nonumber \\
&&\frac{2\pi \beta \delta (k_{1}-k_{1}^{\prime })}{k_{1}^{2}/2+k_{2}^{\prime
2}/2-E}-\beta \int \frac{dk_{2}}{2\pi }\frac{D(k_{2}|k_{1}^{\prime
},k_{2}^{\prime };E)}{k_{1}^{2}/2+k_{2}^{2}/2-E}.  \label{eq:14}
\end{eqnarray}
We are interested in $W=\alpha ^{3}(2\pi )^{-2}\int \int
dkdpD(k|-p,p;E)(k^{2}-E)^{-1}.$ Integrating Eq. (\ref{eq:14}) with respect
to $k_{2}^{\prime }=-k_{1}^{\prime }\equiv -p$ one obtains for $B(k;E)\equiv
(2\pi )^{-1}\int dpD(k|-p,p;E)$ the following equation: 
\begin{eqnarray}
B(k; &E&)\left[ 1+\frac{\beta }{(k^{2}-2E)^{1/2}}\right]   \nonumber \\
&+&\beta \int \frac{dk^{\prime }}{2\pi }\frac{B(k^{\prime };E)}{%
k^{2}/2+k^{\prime 2}/2-E}=\frac{\beta }{k^{2}-E}.  \label{fifteen}
\end{eqnarray}
It has the solution 
\begin{equation}
B(k;E)=\frac{\beta }{\left( k^{2}-E\right) \left( 1+\beta /\sqrt{-E}\right) }%
,  \label{fifteenone}
\end{equation}
which is verified by direct substitution into Eq.~(\ref{fifteen}). Finally
we obtain 
\begin{equation}
W=\alpha ^{3}\int \frac{dk}{2\pi }\frac{B(k;E)}{k^{2}-E}=\frac{2\alpha \beta 
}{\alpha +2\beta }.  \label{sixteen}
\end{equation}
This result together with Eq.~(\ref{eleven}) solves the problem of the
elastic scattering of slow bound pairs for any strength of the short-range
attractive and scattering potentials. It is instructive to compare the pair
transmission ${\cal T}_{2}(Q)$, Eq.~(\ref{eleven}), with the single electron
transmission ${\cal T}_{1}(p)=p^{2}/(p^{2}+\beta ^{2})$ for equal kinetic
energies $p^{2}/2=Q^{2}/4\equiv K$. If the binding potential is strong
compared with the scattering potential ($\alpha \gg 2\beta $) the pair
transmission is just the single-particle transmission of a particle with a
double mass and double barrier strength, ${\cal T}_{2}(Q)=Q^{2}/(Q^{2}+16%
\beta ^{2})$, in accordance with a naive expectation. In the general case
the ratio is 
\begin{equation}
\frac{{\cal T}_{2}(Q)}{{\cal T}_{1}(p)}=\frac{K+\beta ^{2}}{K+4\beta
^{2}(1+2\beta /\alpha )^{-2}}.  \label{seventeen}
\end{equation}
When the binding potential is weaker than the scattering potential ($\alpha
\ll \beta $) the ratio is 
\begin{equation}
\frac{{\cal T}_{2}(Q)}{{\cal T}_{1}(p)}=\left( \frac{\beta }{\alpha }\right)
^{2}\gg 1.  \label{eighteen}
\end{equation}
Quite remarkably, a weak attraction between carriers helps the first
transmitted particle to ``pull'' its partner through a strong potential
barrier.

Another important difference between pair and single-electron tunnelling
occurs due to their different statistics. While electrons are fermions,
preformed pairs are bosons, so that their center-of-mass motion obeys the
Bose-Einstein statistics \ Hence, tunnelling conductance should be
temperature dependent even at low temperatures $T$ as has been already
established in the bipolaron tunnelling to a normal metal with a decay of
the bound state \cite{alekaz}. Here we calculate the current-voltage
characteristics of a molecular junction (MBM), i.e. the current through a
thin potential barrier between two molecular wires. For simplicity, we
restrict our calculations to the strong-coupling \ regime, $\alpha \gg \beta
,T^{1/2}.$ In this regime single carriers are frozen out, and the
transmission is due to the pairs alone,\ which are scattered off a
double-strength barrier, $W\approx 2\beta $, Eq.(\ref{sixteen}), analogously
to single particles with the double carrier mass. Then, in the presence of a
voltage drop at the junction, $2eV$ (for a pair), the conductance can be
readily found by matching the center-of-mass wave function and its
derivative on the left, $\Omega _{l}$ and on the right side, $\Omega _{r}$
of the $\delta $-function barrier. In the coordinate representation one has $%
\Omega _{l}(X<0)=e^{iQX}+R\,e^{-iQX}$, and $\Omega _{r}(X>0)=Ce^{iP_{+}X}$
with $1+R=C$, $CP_{+}-(1-R)Q=8i\beta (1+R)$, and $P_{+}=(Q^{2}+8eV)^{1/2}$.
The transmission is given by 
\begin{equation}
{\cal T}_{2}(Q,P_{+})\equiv 1-|R|^{2}=\frac{4QP_{+}}{(Q+P_{+})^{2}+64\beta
^{2}},  \label{twentyone}
\end{equation}
for real $P_{+}$, and is zero otherwise. Multiplying the transmission by $eQ$
and integrating with the Bose-Einstein distribution function $f(Q)=[\exp
(Q^{2}/4T-\mu /T)-1]^{-1}$ yields the current as 
\begin{equation}
I(V)=e\int_{0}^{\infty }\frac{dQ}{2\pi }Qf(Q)[{\cal T}_{2}(Q,P_{+})-{\cal T}%
_{2}(Q,P_{-})],  \label{twentytwo}
\end{equation}
where $P_{-}=(Q^{2}-8eV)^{1/2}$ and $\mu $ is the chemical potential
determined by the number of {\em pairs} $n$ using $\int_{-\infty }^{\infty }%
\frac{dQ}{2\pi }f(Q)=n.$%
\begin{figure}[t]
\begin{center}
\leavevmode
\hbox{
\epsfxsize=8.4cm
\epsffile{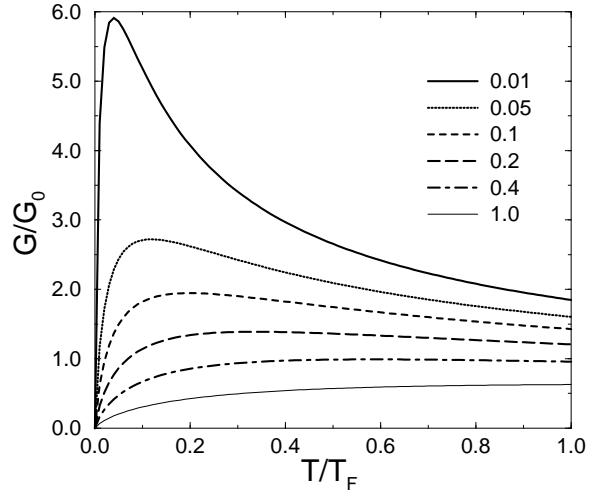}
}
\end{center}
\vspace{-0.5cm}
\caption{ Zero-voltage conductance of MBM as a function of temperature (in
units of $T_{F}$) for different relative strength of the barrier $4\protect%
\beta ^{2}/T_{F}$. $G_{0}=(2e^{2})/h$. }
\label{fig1}
\end{figure}
It is easy to calculate the integrals in the linear voltage classical limit, 
$2eV,T_{F}\ll T$ by expanding the transmission in powers of $eV$ and
replacing the Bose-Einstein distribution with the Boltzmann one, $%
f(Q)\approx (2T_{F}/\pi T)^{1/2}\exp (-Q^{2}/4T)$ ($T_{F}\equiv \pi
^{2}n^{2}/2$ is the Fermi temperature of single carriers). The result for
the conductance, $\sigma \equiv \left( dI/dV\right) _{V=0}$ is 
\begin{equation}
\sigma =\frac{2e^{2}}{\pi }\sqrt{\frac{2T_{F}}{\pi T}}\left[ 1+\frac{4\beta
^{2}}{T}e^{4\beta ^{2}/T}{\rm Ei}(-4\beta ^{2}/T)\right] ,
\label{twentyfour}
\end{equation}
where ${\rm Ei}(x)$ is the exponential integral function. The conductance
behaves as $\sigma =\frac{e^{2}}{\pi \beta ^{2}}\sqrt{\frac{T_{F}T}{2\pi }}$
at $T\ll 4\beta ^{2}$, and as $\sigma =\frac{2e^{2}}{\pi }\sqrt{\frac{2T_{F}%
}{\pi T}}$ at $T\gg 4\beta ^{2}$. In the last case it has a universal
magnitude independent of the barrier strength. Apart from numerical
coefficients, conductance of tightly bound pairs is, of course, the same as
conductance of single electrons in the classical limit. It is not the case,
however, in a degenerate system, when $T\leq T_{F}$. Numerical integration
of Eq.(\ref{twentytwo}) at fixed density $n$ reveals a temperature
dependence in this limit, Fig.~\ref{fig1}, in comparison with the
temperature independent conductance of fermionic noninteracting carriers at
low temperatures. This remarkable difference is entirely due to the bosonic
nature of pairs. The conductance is proportional to the mean velocity of
carriers which in the case of bosons grows as $\sqrt{T}$ (while it is
temperature-independent for fermions). This explains the low-temperature
behavior of the conductance. Interpair correlations may reduce the
difference in 1D wires. However, higher-dimension corrections readily
restore it. There is also a breakdown of Ohm's law when $2eV\geq T$, as
shown in Fig.~\ref{fig2} for low temperatures, again in contrast with the
Fermi statistics, where a non-linearity appears only at $eV\geq T_{F}\gg T$.
\ We suggest that the most appropriate materials for experimental
observation of the unusual current-voltage characteristics (Figs.1,2) are
doped molecular semiconductors such as Cs-doped biphenyl \cite{ramsey90},\
where bipolarons were explicitely detected by photoelectron and
electron-energy-loss spectroscopies, and single crystals of pentacene,
tetracene, rubrene, quaterthiophene ($\alpha $-4T), sexithiophene ($\alpha $%
-8T), where the coherent (bi)polaron tunnelling has been recently observed
below room temperature \cite{bat}.

\begin{figure}[t]
\begin{center}
\leavevmode
\hbox{
\epsfxsize=8.4cm
\epsffile{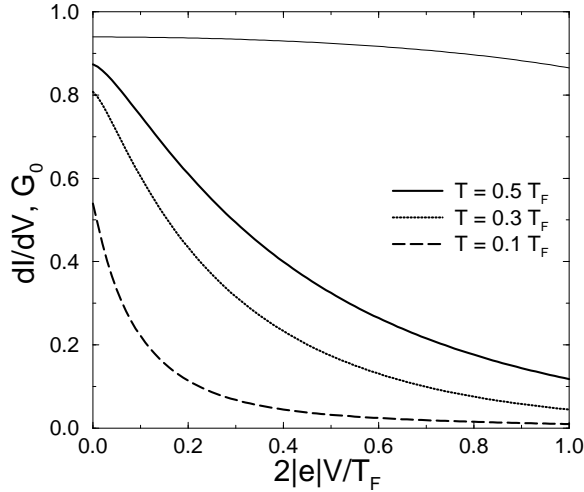}
}
\end{center}
\vspace{-0.5cm}
\caption{ Differential conductance of MBM as a function of voltage for
different temperatures and $4\protect\beta^{2}/T_{F} = 0.5$. The thin solid
lines is the conductance of fermions at $T=0.1\, T_F$. }
\label{fig2}
\end{figure}

In conclusion, we have solved the Lippmann-Schwinger equation in the
effective mass approximation for single carriers and pairs in 1D conductors
(molecular wires), which is valid for (bi)polarons if their size is larger
than the lattice constant. While mapping of this problem onto discrete
lattices is straightforward with a negative Hubbard $U$ model, the model
itself can be applied to bipolarons only in the extreme nonadiabatic limit
when the characteristic phonon frequency is larger than the binding energy 
\cite{alemot}. In this limit a discrete Lippmann-Schwinger equation also has
the analytical solution for slow pairs \cite{alebrakor}, which shows that
the continuous model remains qualitatively correct even \ for lattice size
(small) nonadiabatic bipolarons. In the opposite adiabatic regime bipolaron
tunnelling is not a three-body problem because of the emission and
absorption of (virtual) phonons \cite{alemot}. We have found the scattering
amplitude of elastic scattering of slow bipolarons, and conductance of the
molecular junction (MBM) with preformed pairs. While the pair transmission
is smaller than the single-electron transmission in the strong-coupling
regime, it is surprisingly larger in the weak coupling regime. The
current-voltage characteristics of MBM junction show unusual temperature and
non-linear voltage behavior, Figs.\ref{fig1},\ref{fig2}.

We acknowledge interesting discussions with J.P. Keating, V.V. Osipov, and
R.S. Williams.

\end{document}